# Manufacture and Characterization of Graphene Membranes with Suspended Silicon Proof Masses for MEMS and NEMS Applications

Xuge Fan[1*], Anderson D. Smith[3], Fredrik Forsberg[1], Stefan Wagner[2], Stephan Schröder[1], Sayedeh Shirin Afyouni Akbari[4], Andreas C. Fischer[1, 5], Luis Guillermo Villanueva[4], Mikael Östling[3], Max C. Lemme[2, 3*] and Frank Niklaus[1*]

[1]Division of Micro and Nanosystems, School of Electrical Engineering and Computer Science, KTH Royal Institute of Technology, SE-10044 Stockholm, Sweden.

[2]Faculty of Electrical Engineering and Information Technology, RWTH Aachen University, Otto-Blumenthal-Str. 25, 52074 Aachen, Germany.

[3]Division of Integrated Devices and Circuits, School of Electrical Engineering and Computer Science, KTH Royal Institute of Technology, SE-164 40 Kista, Sweden.

[4]Advanced NEMS Group, École Polytechnique Fédérale de Lausanne (EPFL), 1015 Lausanne, Switzerland.

[5]Silex Microsystems AB, 175 26 Järfälla, Sweden.

Corresponding authors

Name: Xuge Fan, *E-mail: (X.F.) xuge@eecs.kth.se

Name: Frank Niklaus, *E-mail: (F.N.) frank.niklaus@eecs.kth.se

Name: Max C. Lemme, *E-mail: (M.C.L.) lemme@amo.de

**Abstract**

Graphene's unparalleled strength, chemical stability, ultimate surface-to-volume ratio and excellent electronic properties make it an ideal candidate as a material for membranes in micro- and nanoelectromechanical systems (MEMS and NEMS). However, the integration of graphene into MEMS or NEMS devices and suspended structures such as proof masses on graphene membranes raises several technological challenges, including collapse and rupture of the graphene. We have developed a robust route for realizing membranes made of double-layer CVD graphene and suspending large silicon proof masses on membranes with high yields. We have demonstrated the manufacture of square graphene membranes with side lengths from 7 μm to 110 μm and suspended proof masses consisting of solid silicon cubes that are from 5 μm × 5 μm × 16.4 μm to 100 μm × 100 μm × 16.4 μm in size. Our approach is compatible with wafer-scale MEMS and semiconductor manufacturing technologies, and the manufacturing yields of the graphene membranes with suspended proof masses were greater than 90%, with more than 70% of the graphene membranes having more than 90% graphene area without visible defects. The measured resonance frequencies of the realized structures ranged from tens to hundreds of kHz, with quality factors ranging from 63 to 148. The graphene membranes with suspended proof masses were extremely robust and were able to withstand indentation forces from an atomic force microscope (AFM) tip of up to ~7000 nN. The proposed approach for the reliable and large-scale manufacture of graphene membranes with suspended proof masses will enable the development and study of innovative NEMS devices with new functionalities and improved performances.

# Introduction

The atomically thin structure of graphene (atom-layer distance of ~0.335 nm) and its remarkable mechanical[1] and electrical properties[2] (Young's modulus of up to ~1 TPa and charge carrier mobility of up to 200,000 cm$^2$ V$^{-1}$ s$^{-1}$) make it a very promising membrane and transducer material for micro- and nanoelectromechanical system (MEMS & NEMS) applications[3–9]. However, the application of suspended graphene in NEMS devices has thus far been limited to resonators[10–19], pressure sensors[20–25], switches[7,26–28], loudspeakers[29], microphones[30,31] and devices for fundamental studies of the material and structural properties of graphene[8,32–37]. The reported suspended graphene structures include doubly clamped graphene beams, fully clamped graphene drums and suspended graphene-based cantilevers. Suspended structures are typically realized by transferring graphene from the original substrate to a pre-fabricated substrate with trenches[11,15], cavities[12,20] or membranes made of dielectric layers[21,38,38,39] or by transferring graphene from the original substrate to a flat silicon dioxide (SiO$_2$)[16,17,40–42] or polymer substrate surface[43,44] and then removing parts of the material underneath the graphene by sacrificial etching.

In contrast to previously reported graphene membranes and beams, MEMS and NEMS devices such as accelerometers, gyroscopes and resonators often employ larger proof masses (e.g., ~$10^7$ to ~$10^{10}$ μm$^3$ in size) that are suspended on springs in the form of membranes, beams or cantilevers. Graphene, as a robust and intrinsically nanoscale material, could be used to suspend large proof masses, thereby forming spring-mass systems consisting of atomically thin graphene springs for potential applications as ultra-miniaturized transducer elements in future high-performance NEMS devices [20]. However, the realization of suspended graphene with large attached proof masses is difficult, and to the best of our knowledge, no such examples have been reported in the literature.

A previous report of suspended graphene membranes with very small masses included micrometre-sized few-layer graphene cantilevers with diamond allotrope carbon masses (0.5 μm in length, 1.5 μm in width and 20 nm in thickness, with a corresponding weight of $5.7 \times 10^{-14}$ g) fabricated using focused ion beam (FIB) deposition for the study of the mechanical properties of graphene[45]. Previous literature also reports a spiral spring, a kirigami pyramid and a variety of cantilevers based on a suspended graphene monolayer supporting 50 nm thick gold masses attached to suspended cantilevers[46]. However, these structures had to be kept in a liquid to maintain their mechanical integrity. Suspended graphene membranes with diameters of 3-10 μm that were circularly clamped by a polymer (SU-8) and that supported a mass made of either SU-8 or gold located at the centre of the membrane were reported for shock detection caused by ultra-high mechanical impacts[47]. However, all previous reports involved extremely small masses, and the fabrication methods employed, such as FIB-induced deposition, were slow and typically not compatible with large-scale manufacturing.

In this paper, we present a robust, scalable and high-yield manufacturing approach to realize CVD graphene membranes with large suspended silicon (Si) proof masses that is compatible with MEMS and NEMS manufacturing processes and that can be utilized for devising NEMS with graphene membranes as structural and functional components. Our approach employs a silicon-on-insulator (SOI) substrate to form silicon proof masses that are etched in the silicon device layer of the SOI wafer. The graphene membranes are formed by transferring a double layer of CVD graphene to the pre-patterned SOI wafer, followed by a combination of dry etching and vapour HF etching of the buried oxide (BOX) layer to release the silicon proof masses and suspend them on the graphene membranes. Static and dynamic mechanical characterization of the manufactured structures shows that they are robust and can potentially be used as spring-mass systems in future ultra-small NEMS such as resonators and accelerometers.

**Results**

To demonstrate the feasibility of graphene membranes with large suspended proof masses, we fabricated square membranes with different dimensions made of double-layer graphene on which silicon proof masses of different sizes were suspended. A typical device structure is illustrated in Fig. 1a-e. Our fabrication approach utilizes an SOI wafer where the silicon proof mass is formed in the device layer of the SOI wafer by dry etching, followed by transfer of double-layer graphene to the SOI wafer and release of the proof mass by sacrificially removing the BOX layer using dry etching in combination with vapour HF etching. A schematic of the fabrication and integration process is shown in Fig. 1f-i, and three-dimensional (3D) and cross-sectional views of the structure at key process steps are shown in Fig. 2 (see Methods section for details). Scanning electron microscopy (SEM) images of typical graphene membrane structures with suspended silicon proof masses are shown in Fig. 3. The process scheme of patterning the SOI substrate prior to graphene transfer reduces the processing steps after graphene transfer, which improves the cleanliness and purity of the graphene and reduces the risk of rupturing or destroying the membranes during processing. Each layer of the SOI substrate has a specialized function: the device layer is used for fabricating trenches and defining the proof masses, the handle substrate is used as a support, and the BOX layer is used as a sacrificial layer to gently release the mass. In our demonstration, we have chosen SOI wafers with thicknesses of the silicon device layer, the BOX layer and the handle layer of 15 μm, 2 μm, and 400 μm, respectively, as depicted in Fig. 1c. Here, the thickness of the silicon device layer was chosen to form proof masses with reasonable aspect ratios but can in principle be adapted to specific application requirements.

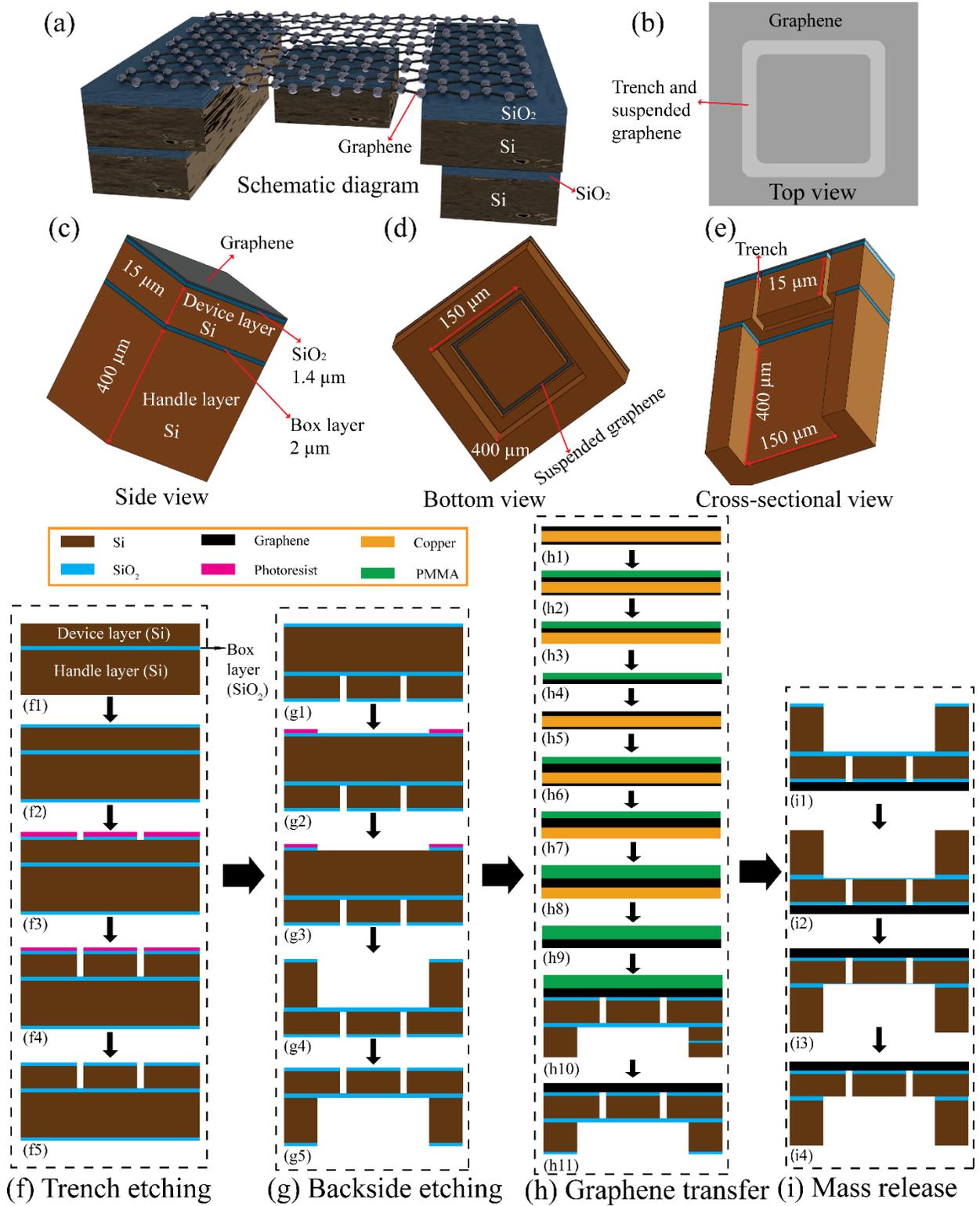

**Fig. 1 3D diagrams of the structures and SEM images**. **a** 3D schematic of the graphene membrane with a suspended proof mass. **b, c, d, e** 3D schematic top view, side view, bottom view

and cross-sectional view, respectively. Schematic of the fabrication and integration process: **f** Trench etching: (f1) SOI wafer, (f2) oxidation of both wafer sides, (f3) trench etching of the $SiO_2$ layer on the silicon device layer, (f4) trench etching of the silicon device layer, (f5) removal of PR residues. **g** Backside etching: (g1) backside of the chip, (g2) patterning of the PR layer on the backside of the chip, (g3) backside etching of the $SiO_2$ layer, (g4) backside etching of the handle substrate, (g5) the chip after backside etching. **h** Graphene transfer: (h1) monolayer graphene on a copper sheet, (h2) spin coating of PMMA, (h3) etching of carbon residues on the backside of the copper sheet, (h4) dissolution of the copper in $FeCl_3$, (h5) graphene monolayer on a second copper sheet, (h6) transfer of the PMMA/graphene stack to the graphene on the second copper sheet, (h7) etching of the carbon residues from the backside of the copper sheet, (h8) spinning of PMMA on the graphene, (h9) dissolution of the copper in $FeCl_3$, (h10) transfer of the double-layer graphene stack on the pre-patterned SOI substrate. **i** Proof mass release: (i1) backside of the chip, (i2) RIE etching of the BOX layer until a thin (~ 100 nm) $SiO_2$ layer remains, (i3) the chip after RIE etching, (i4) vapour HF etching to remove the remaining thin $SiO_2$ layer.

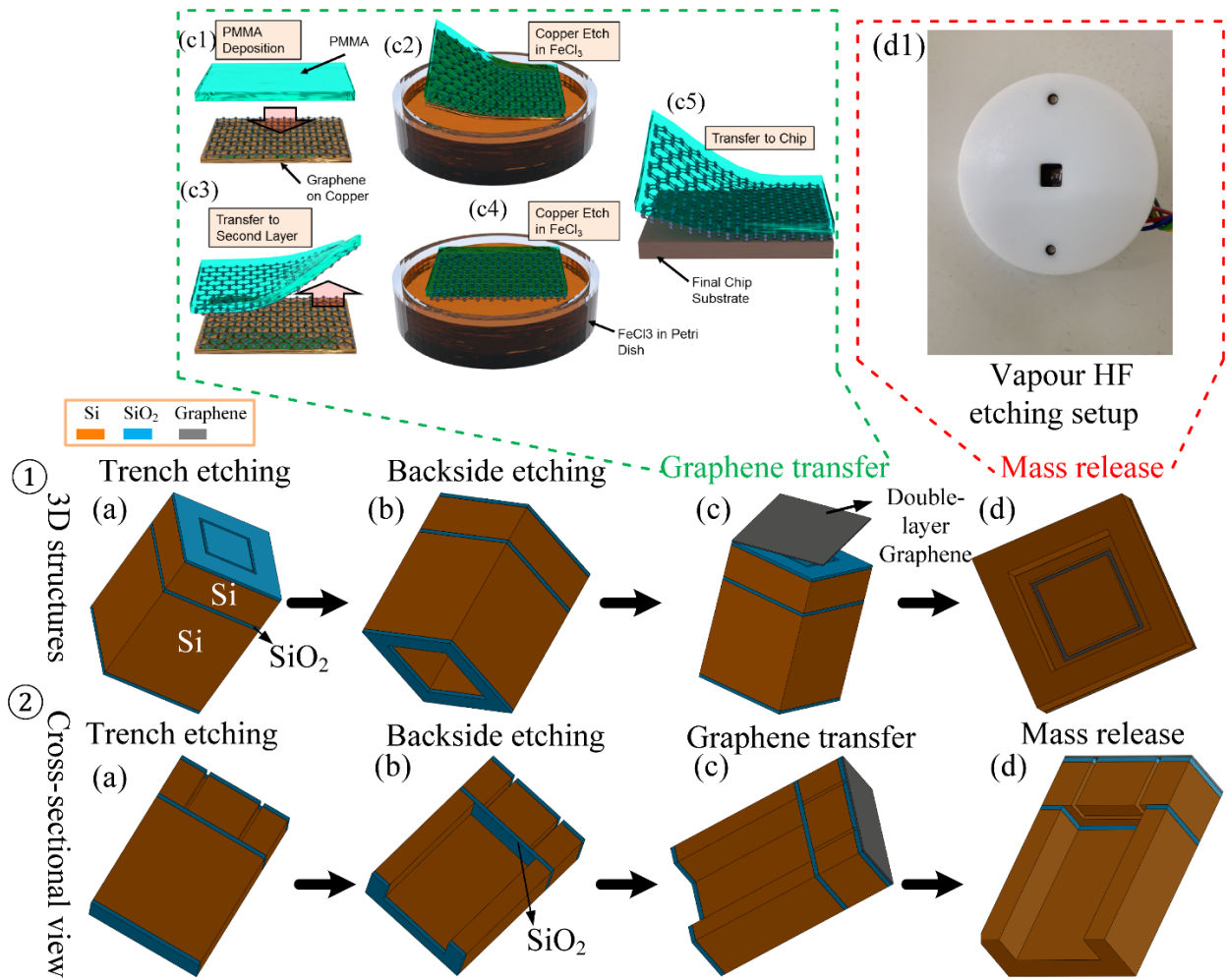

**Fig. 2 Schematic of key fabrication and integration process steps in 3D (①) and cross-sectional (②) views**. **a** Trench etching. **b** Backside etching. **c** Details of the graphene transfer. **d** Mass release by dry etching followed by vapour HF etching.

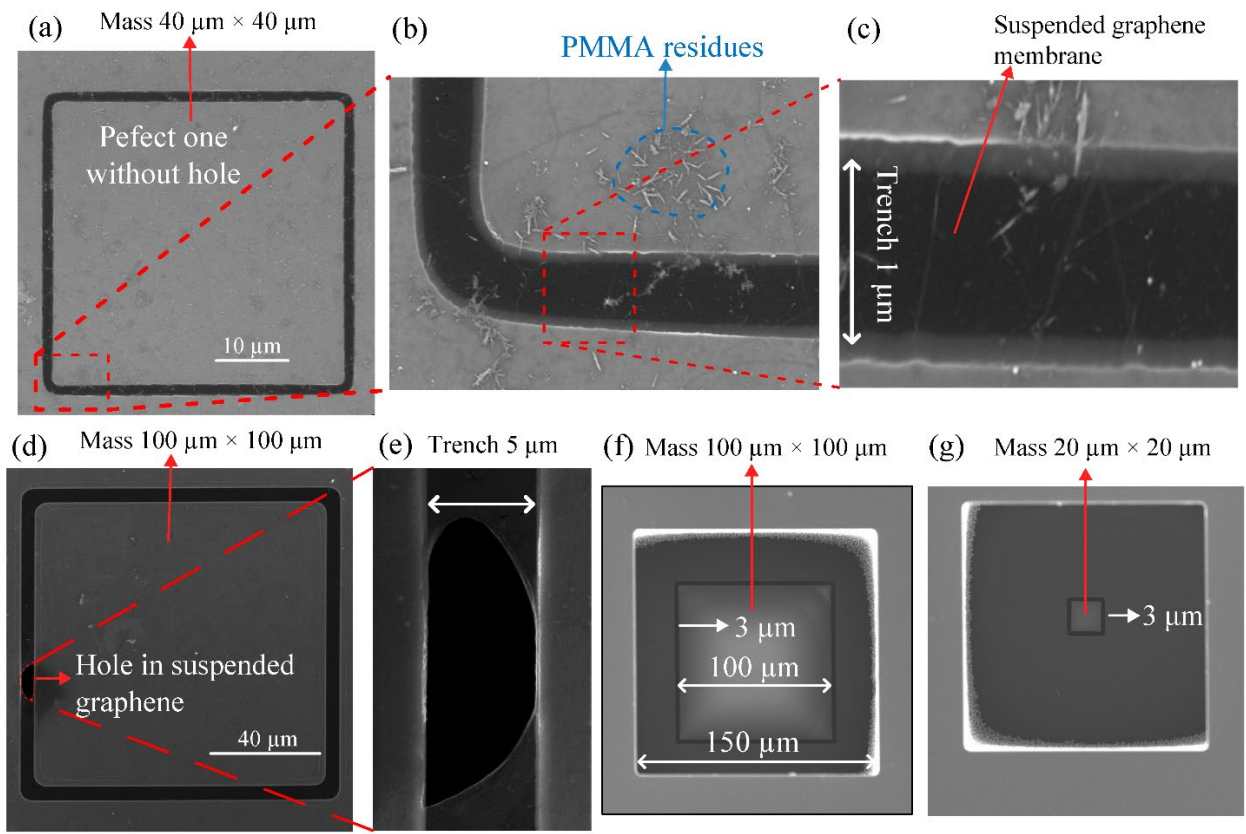

**Fig. 3** SEM characterizations of graphene membranes with suspended proof masses. **a, b, c** SEM images of the top side of a structure with a 1 μm wide trench and 40 μm × 40 μm × 16.4 μm sized proof mass. **d, e** SEM images of the top side of a structure with a 5 μm wide trench and a 100 μm × 100 μm × 16.4 μm sized proof mass. **f, g** SEM images of the bottom side of a structure with a 3 μm wide trench and 100 μm × 100 μm × 16.4 μm and 20 μm × 20 μm × 16.4 μm sized proof masses, respectively.

To demonstrate the flexibility and robustness of our fabrication process and of the resulting graphene structures, we designed and fabricated structures with different trench widths and the silicon proof mass dimensions. The dimensions of the smallest trenches were 1 μm × 7 μm, and the dimensions of the largest trenches were 5 μm × 110 μm, resulting in square graphene membranes with dimensions from 7 μm × 7 μm to 110 μm × 110 μm. The smallest proof mass suspended on a graphene membrane consisted of a square cuboid measuring 5 μm × 5 μm × 16.4 μm, and the largest mass consisted of a square cuboid measuring 100 μm × 100 μm × 16.4 μm. The trench depth was 16.4 μm, which is identical to the thickness of the silicon mass, consisting of a 15 μm thick silicon device layer and a 1.4 μm thick $SiO_2$ layer (Fig. 1c and e). The calculated weight of a 100 μm × 100 μm × 15 μm silicon proof mass covered by a 1.4 μm thick layer of $SiO_2$ is $3.86 \times 10^{-7}$ g, where the $SiO_2$ and the silicon densities are $2.65 \times 10^3$ kg/m$^3$ and $2.329 \times 10^3$ kg/m$^3$, respectively. The side length and depth of the open space formed by backside etching of the handle substrate of the SOI wafer are 150 μm × 150 μm and 400 μm, respectively (Fig. 1d and e). From Table 1, it can be seen that the weight of this proof mass is three orders of magnitude larger than the SU-8 mass, six orders of magnitude larger than the gold mass and seven orders of magnitude larger than the carbon mass that have been reported previously, respectively[45–47]. The dramatically increased weight of the suspended proof mass is potentially of interest for applications such as miniaturized NEMS inertial sensors.

SEM images of different graphene membranes with suspended proof masses are shown in Fig. 3a-c and Fig. 3d and e, with the structure in Fig. 3a-c being free of holes in the suspended graphene membrane and the structure in Fig. 3d and e featuring a hole in the graphene membrane. To demonstrate the complete release of the proof mass, SEM characterization of the backside of the SOI chips was performed (Fig. 3 f and g). From these SEM images, it can be seen that the BOX layer was completely removed, which means that our fabricated proof masses are suspended and only attached to the graphene membranes. The strong attachment of the $SiO_2$/Si proof mass to the graphene membrane is due to the large adhesion energy between the graphene and the $SiO_2$ surface of the proof mass that is caused by van der Waals forces[48,49]. Extremely strong adhesion of graphene to SiO$_2$ surfaces by van der Waals interactions has been previously demonstrated by experiments, analytical models and atomistic simulations[49,50].

To verify that double-layer graphene indeed exists in our fabricated structures, we performed Raman spectroscopy. Fig. 4a shows the Raman spectra of double-layer graphene at three different positions (Fig. 4b) of a manufactured structure with 4 μm wide trenches and a 50 μm × 50 μm × 16.4 μm proof mass. The Raman spectra show the typical characteristic peaks of graphene: The "G peaks" at approximately 1600 cm$^{-1}$ (Fig. 4c) and the "2D peaks" at approximately 2700 cm$^{-1}$ (Fig. 4d) [51,52] demonstrate the presence of graphene, and the absence of an appreciable D peak (1350 cm$^{-1}$) in the Raman spectra indicates the relatively high quality of the graphene. The shape of the 2D band in Fig. 4d indicates the presence of double-layer graphene. The second-order Raman 2D band caused by a two-phonon lattice vibrational process is sensitive to the number of layers of graphene, and the 2D band of monolayer graphene is very sharp and symmetric[53]. For double-layer and multi-layer graphene, the 2D band becomes much broader, as shown in Fig. 4, mainly due to the change in the electronic structure of the graphene[53].

**Table 1 Comparison of small masses suspended on graphene membranes reported in the literature.**

| *Material of mass* | *Mass shape* | *Side length* | *Thickness (μm)* | *Area (μm²)* | *Volume (μm³)* | *Density (g/cm³)* | *Weight (g)* |
|---|---|---|---|---|---|---|---|
| Carbon[45] | cube | 0.5 μm × 0.5 μm | 0.02 | 0.75 | 0.015 | 3.8 | 5.7 × 10$^{-14}$ |
| Gold[46] | cube | 10 μm × 10 μm | 0.05 | 100 | 5 | 19.3 | 9.65 × 10$^{-13}$ |
| SU-8[47] | cylinder | Diameter: 10 μm | 1.5 | 78.5 | 117.75 | 1.199 | 1.412 × 10$^{-10}$ |
| Si/SiO$_2$ (This work) | cube | 100 μm × 100 μm | SiO$_2$: 1.4 | 10000 | 14000 | 2.65 | 3.866 × 10$^{-7}$ |
|  |  |  | Si: 15 |  | 150000 | 2.33 |  |

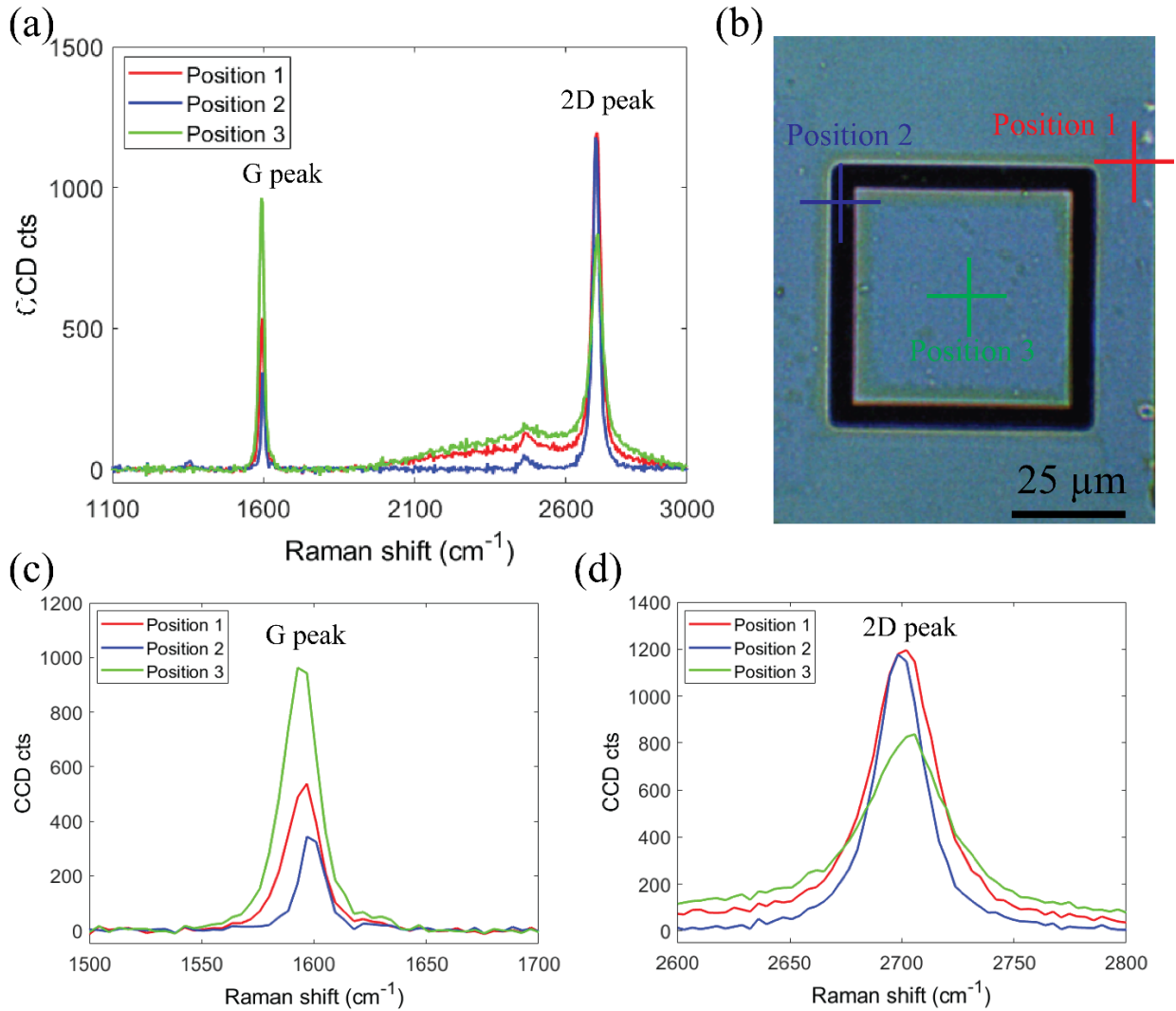

**Fig. 4 Raman spectroscopy of double-layer graphene. a** Raman spectra of the double layer at three different positions of a structure with 4 μm wide trenches and a 50 μm × 50 μm × 16.4 μm proof mass, with "G peaks" at approximately 1596.8 cm$^{-1}$ (position 1), 1596.8 cm$^{-1}$ (position 2) and 1592.6 cm$^{-1}$ (position 3) and "2D peaks" at approximately 2701.9 cm$^{-1}$ (position 1), 2698.3 cm$^{-1}$ (position 2) and 2705.6 cm$^{-1}$ (position 3). **b** Optical microscopy image of the manufactured device in (**a**) at the three different measurement positions. Position 1 (red cross) is on the non-suspended area of double-layer graphene on the substrate; position 2 (blue cross) is on the suspended double-

layer graphene membrane; position 3 (green cross) is on the double-layer graphene on the suspended mass. **c** Magnification of the G peaks in (**a**). **d** Magnification of the 2D peaks in (**a**).

To characterize the dynamic mechanical properties of the spring-mass system of our structures, we determined the resonance frequencies of four structures in vacuum (using a vacuum chamber with $10^{-5}$ mbar actively pumped vacuum) by measuring the amplitude of their thermomechanical noise using laser Doppler vibrometry (LDV) (Fig. 5, Figure S1 and Figure S2 in the Supporting Information, and Methods). Fig. 5a-d shows the LDV measurements of four structures (Fig. 5e-h) that have identical trench widths (3 µm) but different proof mass dimensions (25 µm × 25 µm × 16.4 µm; 30 µm × 30 µm × 16.4 µm; 40 µm × 40 µm × 16.4 µm and 50 µm × 50 µm × 16.4 µm, Fig. 5e-h). The resonance frequencies of these devices are ~158 kHz (Fig. 5a), ~90 kHz (Fig. 5b), ~78.8 kHz (Fig. 5c), and ~60.3 kHz (Fig. 5d). As expected, the resonance frequency decreases with an increase in the weight of the suspended proof mass (Fig. 5a-d). The corresponding quality factor (Q) of one of the structures (Fig. 5g) is estimated by using a Lorentz fitting to their resonance frequencies of approximately 63 (Fig. 5c). The Q factor is comparable to those reported in previous studies[10,11,13,15,40,54]. Since these measurements were performed in vacuum, we can claim that the Q-factors of our structures are likely dominated by energy losses in the mechanical structure itself, such as losses from internal friction in the double-layer graphene membranes, clamping losses, surface losses, and thermoelastic damping[55–57].

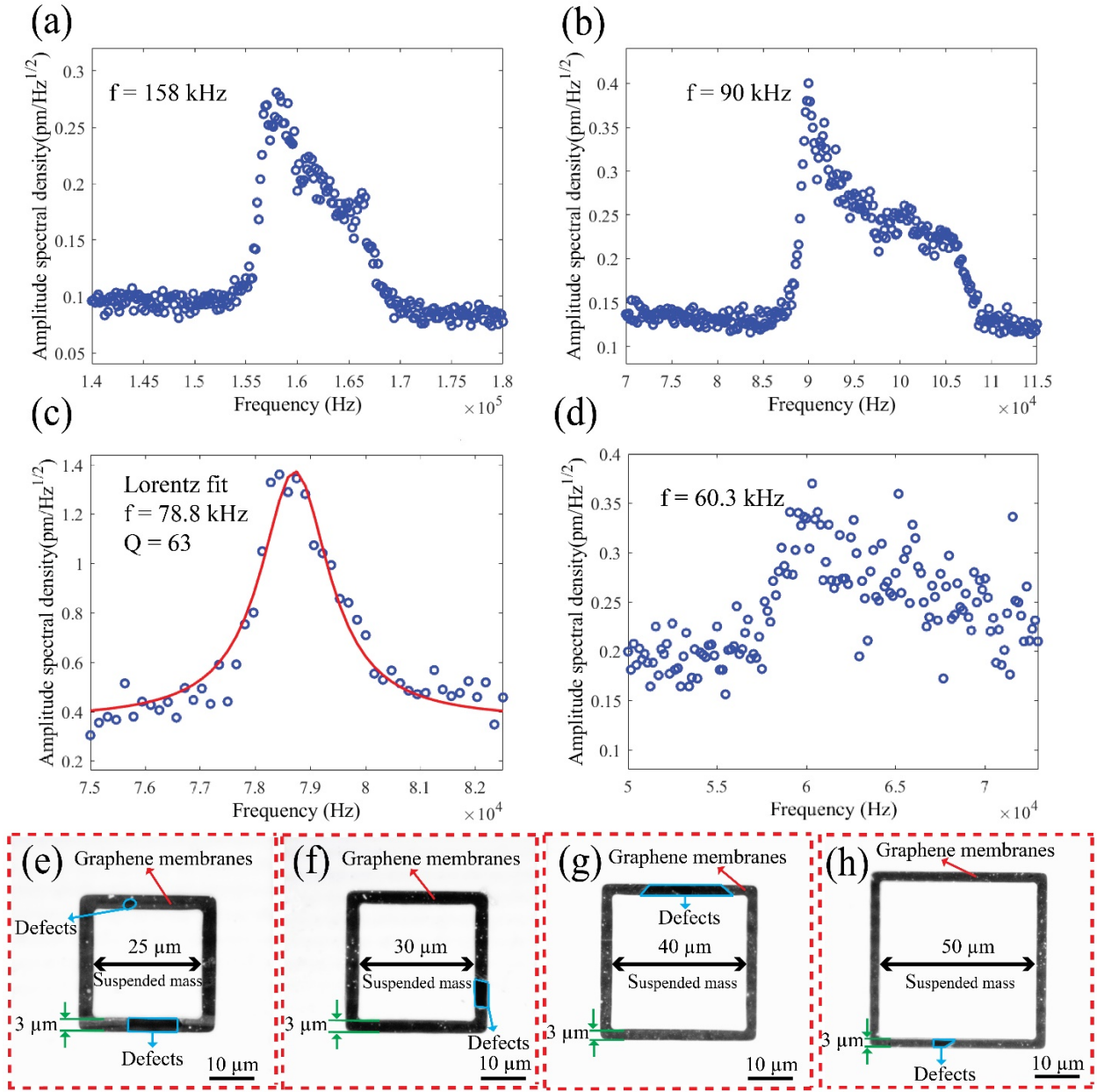

**Fig. 5 Dynamic mechanical characterization of suspended graphene membranes with attached silicon masses by measuring the amplitude of thermomechanical noise in vacuum using laser Doppler vibrometry (LDV). a, b, c, d** Thermomechanical noise peak of four devices using LDV, with resonance frequencies of 158 kHz (**a**), 90 kHz (**b**), 78.8 kHz (**c**) and 60.3 kHz (**d**) and a quality factor of 63 (**c**). The red solid lines in (**c**) are based on Lorentz fitting. The four devices have identical trench widths (3 μm) but different proof mass dimensions (25 μm × 25 μm × 16.4

μm in (**a**); 30 μm × 30 μm × 16.4 μm in (**b**); 40 μm × 40 μm × 16.4 μm in (**c**) and 50 μm × 50 μm × 16.4 μm in (**d**)). **e, f, g, h** High-contrast microscopy images of suspended graphene membranes with attached proof mass of the four measured devices in (**a**), (**b**), (**c**) and (**d**), respectively. The graphene membranes of the four structures have defects with different dimensions and densities.

To further confirm the frequency response of the spring-mass system of our structures, we used LDV to measure the frequency response of a device with a trench width of 3 μm and proof mass dimensions of 50 μm × 50 μm × 16.4 μm in air (atmospheric pressure) at room temperature by driving the device with a piezoshaker (Fig. 6). Fig. 6a displays the amplitude and phase response as a function of frequency. The resonance frequency (88.1 kHz) of the device is of the same order as those found using thermomechanical noise measurements (Fig. 5). The corresponding Q factor of the structure (Fig. 6c) was estimated by using a Lorentz fit to be approximately 148 (Fig. 6b), which is comparable to those based on thermomechanical noise measurements (Fig. 5). Since our graphene membranes and proof masses are not located close to a surface in the direction of the movement of the membrane, we do not expect squeeze-film damping when the structure is operated in gas at atmospheric pressure. The measured value of Q in air confirms our hypothesis.

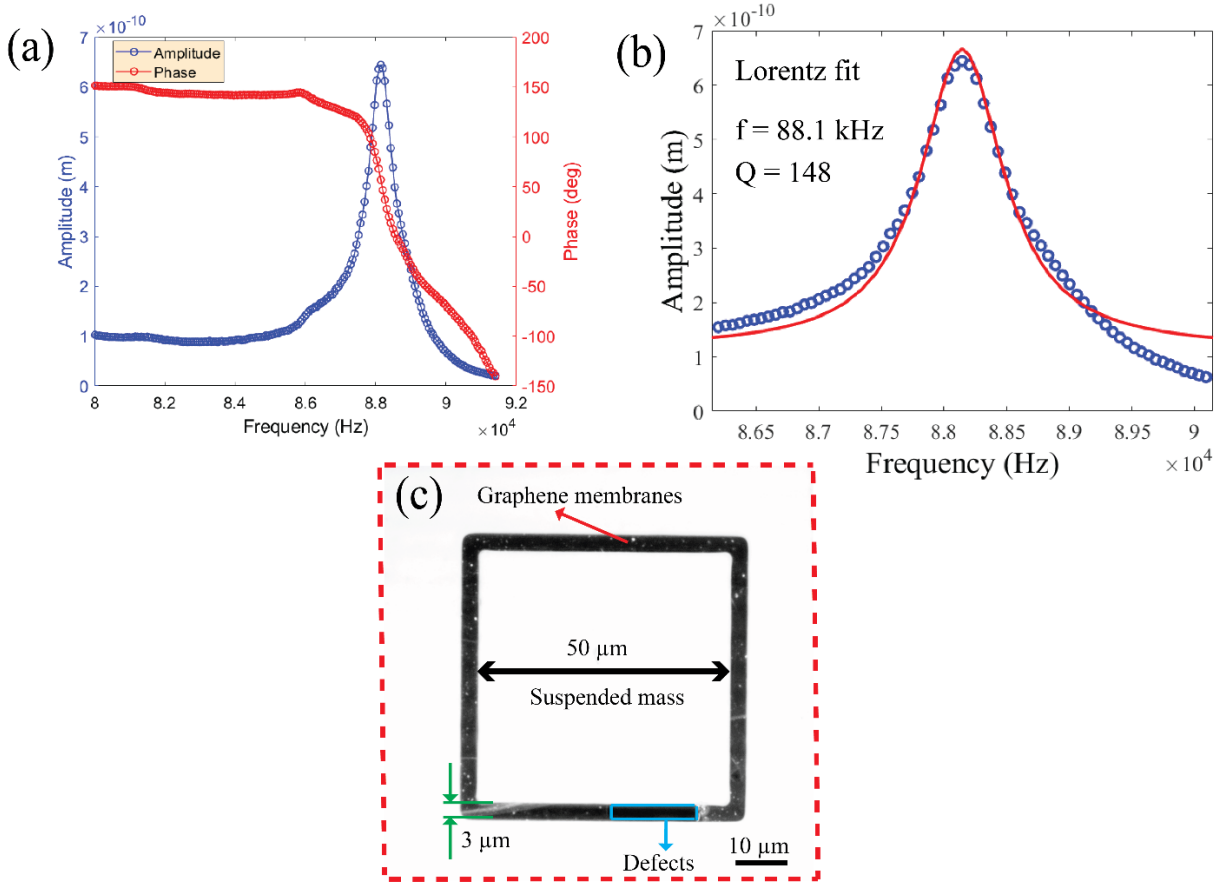

**Fig. 6 Dynamic mechanical characterization using LDV of a device of suspended graphene membranes with an attached silicon mass that was driven by a piezoshaker in air. a** Amplitude (blue line and blue circle marker) and phase (red line and red circle marker) response of a device (trench width: 3 μm: proof mass dimension: 50 μm × 50 μm × 16.4 μm) while performing a frequency scan. **b** Lorentz fitting (red line) of the measured resonant response shown in (**a**). The resonance frequency is 88.1 kHz, and the quality factor is 148. **c** A high-contrast microscopy image of suspended graphene membranes with an attached proof mass of the measured device in (**a**).

To characterize the static mechanical properties and robustness of our graphene structures, we performed force-displacement measurements using AFM tip indentation at the centre of a suspended proof mass of a structure with 4 μm wide trenches and a proof mass size of 20 μm × 20 μm × 16.4 μm (Fig. 7a, b and c). As shown in Fig. 7b, when the AFM indentation force gradually increased from 15.5 nN to 6968 nN, the displacement of the proof mass increased from 7.7 nm to 697 nm. Surprisingly, even when the AFM indention force was increased to 6968 nN, the graphene membrane did not rupture. For reference, the weight of a 20 μm × 20 μm × 16.4 μm large silicon proof mass causes a force due to earth gravity that is on the order of 0.156 nN. Thus, our results illustrate that the suspended graphene membranes with attached proof mass are generally very robust and potentially useful for application in future NEMS inertial sensors. The corresponding average strain in the suspended graphene membranes at the maximum displacement (697 nm) of the proof mass is estimated to be on the same order or smaller than the ones reported in AFM indentation experiments on fully clamped graphene membranes[1,58,59] (Tables S1 and S2 in the Supporting Information). We hypothesize that circular graphene membranes and proof mass designs might have even better mechanical robustness due to the avoidance of corners that are prone to stress concentrations. Another potential advantage of circular membranes and proof mass designs is that circular symmetry may result in more uniform strain distributions in the graphene membrane.

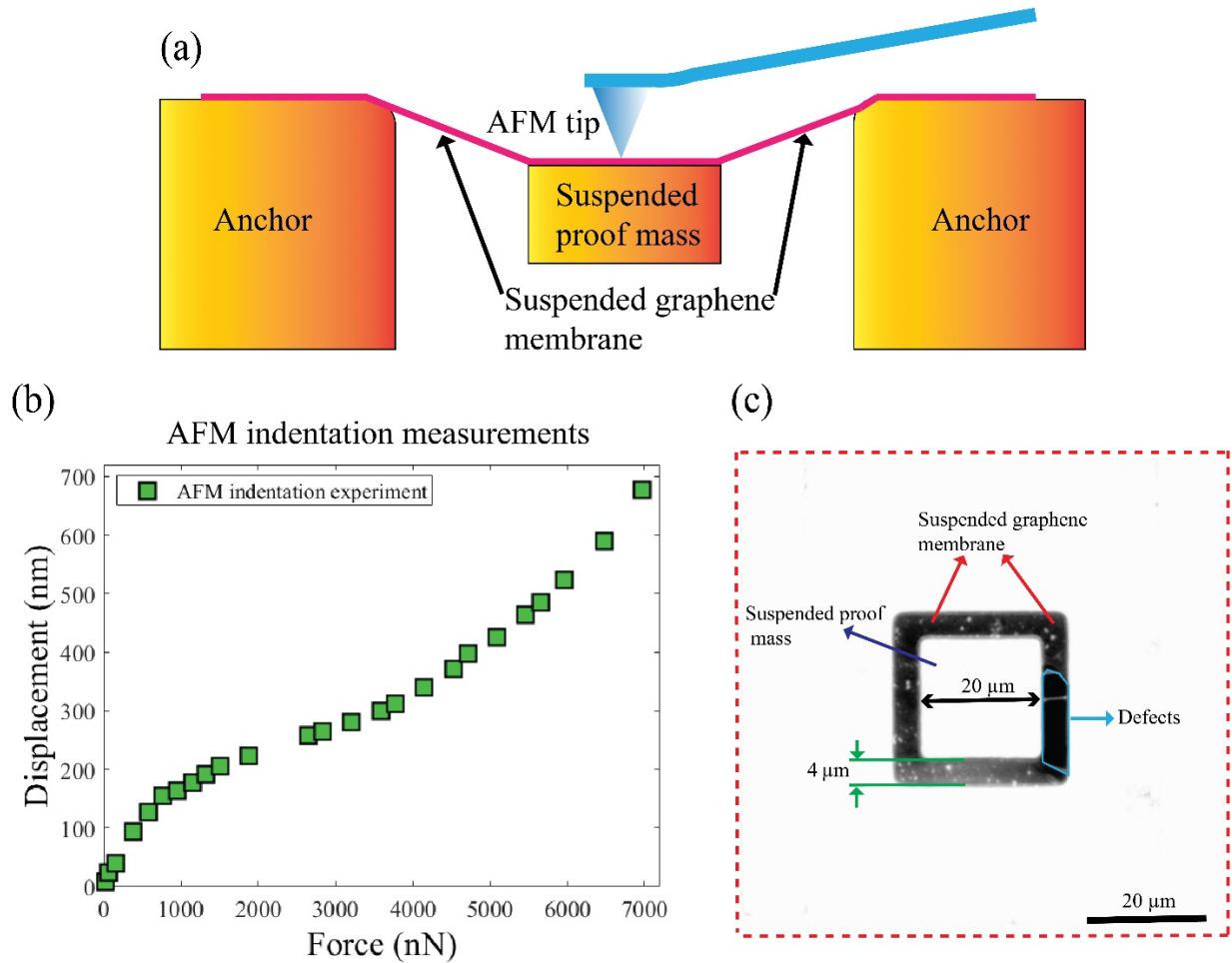

**Fig. 7 Force-displacement measurements of suspended graphene membranes with an attached proof mass by AFM tip indentation.** A Schematic of force-displacement measurement by AFM indentation at the centre of the suspended proof mass. B Force-displacement measurement of a structure with 4 μm wide trenches and a proof mass size of 20 μm × 20 μm × 16.4 μm. c High-contrast microscopy image of the suspended graphene membrane with attached proof mass measured in (**b**).

To analyse the yield of our process, we performed systematic experiments by manufacturing a series of graphene membrane structures with different dimensions (trenches widths from 1 μm to 5 μm and proof masses measuring from 5 μm × 5 μm × 16.4 μm to 100 μm × 100 μm × 16.4 μm) and characterizing the resulting structures by SEM. We fabricated twelve chips that comprised 672 structures in different batches, and we obtained similar yields for all chips. Figs. 8-12 show typical examples of suspended graphene membranes with different trench widths (1 μm, 2 μm, 3 μm, 4 μm, 5 μm) and with attached proof masses of different dimensions (from 5 μm × 5 μm × 16.4 μm to 100 μm × 100 μm × 16.4 μm) after releasing the BOX layers. There are some graphene membrane structures without any holes (Fig. 8a-c and Fig. 11a-b). However, typically, a few small holes in the suspended graphene membranes were present, although most of the structures maintained their mechanical integrity (Fig. 8d, Fig. 9, and Fig. 10). Even for most of the large membranes with wide trenches (5 μm) and large proof masses (100 μm × 100 μm × 16.4 μm), the sizes of the holes were comparably small (Fig. 8d, Fig. 10d, Fig. 11d and Fig. 12). For the small membranes with narrow trenches and small proof masses, slightly more of the obtained graphene membranes lacked holes or had only very small holes. The sizes of the holes in the suspended graphene membranes are related to the dimensions of the attached masses and the widths of the trenches. For instance, for structures with 1 μm wide trenches and 40 μm × 40 μm × 16.4 μm masses, approximately 15% of the structures were defect-free, without any holes in the suspended graphene membranes (Fig. 13a), while approximately 40% of the structures had tiny and small-scale (~0-1 μm in side length) holes (Fig. 13b-c), approximately 35% of the structures had medium-scale (~1-5 μm in side length) holes (Fig. 13d), and approximately 10% of the structures had large-scale (> 5 μm in side length) holes (Fig. 13e). The reasons that such different sizes of holes occurred in the suspended graphene membranes are not presently clear. In-plane tension, shear and compression of the suspended graphene are some possibilities. Another possibility is that during graphene wet transfer, there might be some water remaining in the trenches after transferring graphene. During the subsequent process steps, the water remaining in the trenches might evaporate and rupture the suspended graphene membranes.

In addition, occasional tears might occur at mechanically weak grain boundaries between crystals in the CVD growth graphene. We also estimated the yield in dependence of the area of the suspended graphene membrane over the trenches, and we found that there was no obvious difference among different trench sizes (from 1 μm to 5 μm) or among proof masses with different dimensions (from 5 μm × 5 μm × 16.4 μm to 100 μm × 100 μm × 16.4 μm). In this analysis, we defined the coverage as the percentage of the area of the suspended graphene over the trenches in each structure. For instance, a 100% coverage area, as shown in Fig. 13a, means that there are no holes in the graphene membrane. When the widths of the trenches and the sizes of the masses increase, the size of the holes in the graphene membrane typically increase. However, the ratio of the graphene membrane coverage area to the total trench area was similar among structures with different trenches and mass sizes. In summary, approximately 15% of the graphene membranes had 100% coverage of the trenches, approximately 75% of the graphene membranes had > 90% coverage of the trenches, approximately 90% of the graphene membranes had > 75% coverage of the trenches, and approximately 10% of the graphene membranes had < 75% coverage of the trenches (Fig. 14).

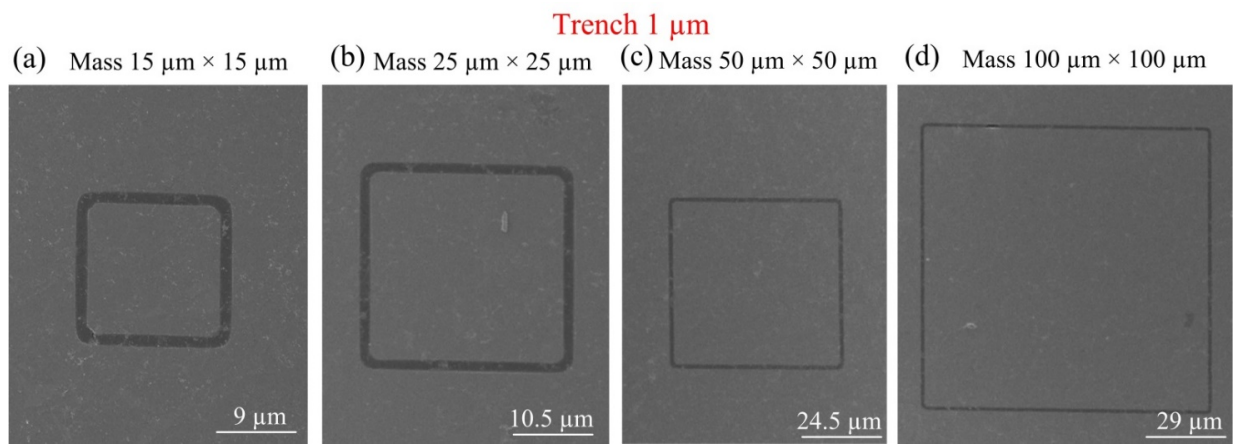

**Fig. 8** SEM images of structures with 1 μm wide trenches and different sizes of proof masses. **a** 15 μm × 15 μm × 16.4 μm mass. **b** 25 μm × 25 μm mass × 16.4 μm. **c** 50 μm × 50 μm × 16.4 μm mass. **d** 100 μm × 100 μm × 16.4 μm mass.

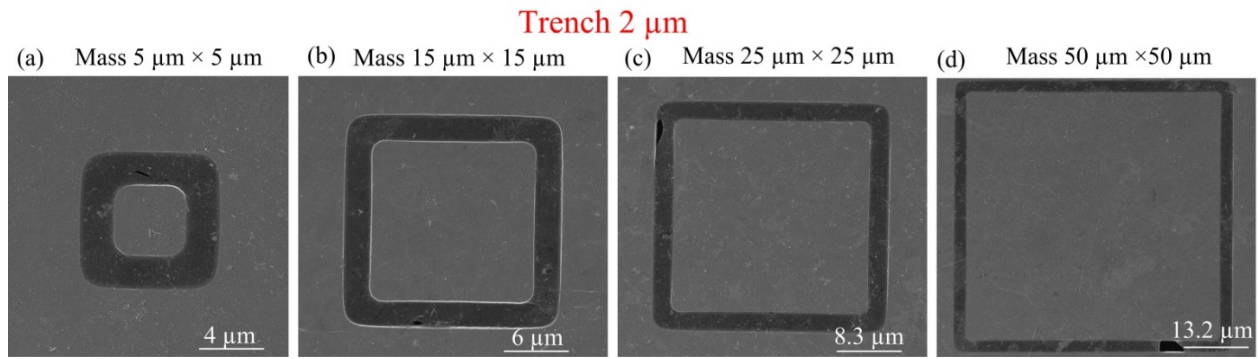

**Fig. 9** SEM images of structures with 2 μm wide trenches and different sizes of proof masses. **a** 5 μm × 5 μm × 16.4 μm mass. **b** 15 μm × 15 μm × 16.4 μm mass. **c** 25 μm × 25 μm × 16.4 μm mass. **d** 50 μm × 50 μm × 16.4 μm mass.

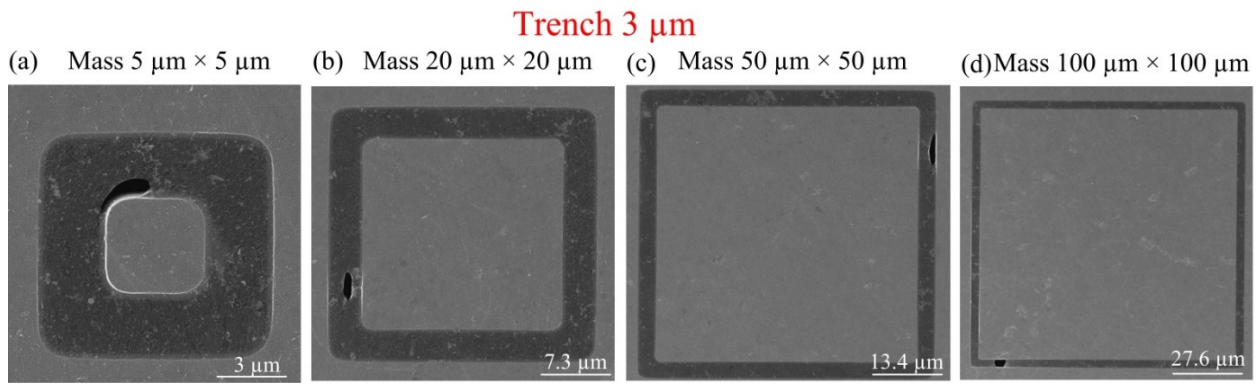

**Fig. 10** SEM images of structures with 3 μm wide trenches and different sizes of proof masses. **a** 5 μm × 5 μm × 16.4 μm mass. **b** 20 μm × 20 μm × 16.4 μm mass. **c** 50 μm × 50 μm × 16.4 μm mass. **d** 100 μm × 100 μm × 16.4 μm mass.

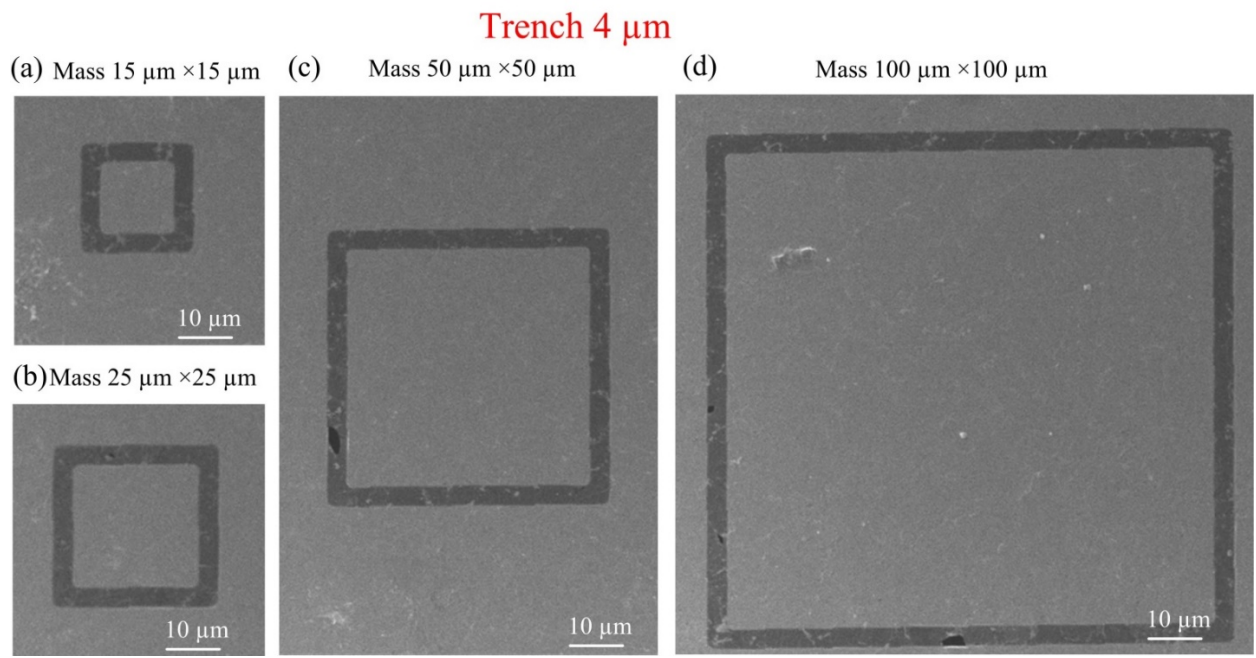

**Fig. 11 SEM images of structures with 4 μm wide trenches and different sizes of proof masses.** **a** 15 μm × 15 μm × 16.4 μm mass. **b** 25 μm × 25 μm × 16.4 μm mass. **c** 50 μm × 50 μm × 16.4 μm mass. **d** 100 μm × 100 μm × 16.4 μm mass.

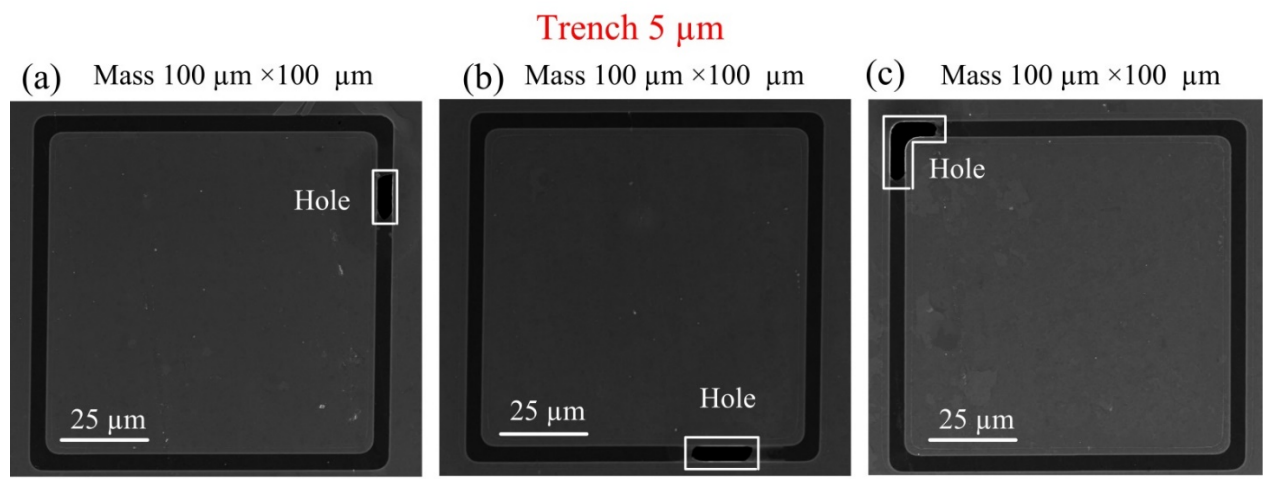

**Fig. 12 SEM images of structures with 5 μm wide trenches and 100 μm × 100 μm × 16.4 μm proof masses.** The white boxes in (**a**), (**b**) and (**c**) label the holes in random positions of suspended double-layer graphene membranes.

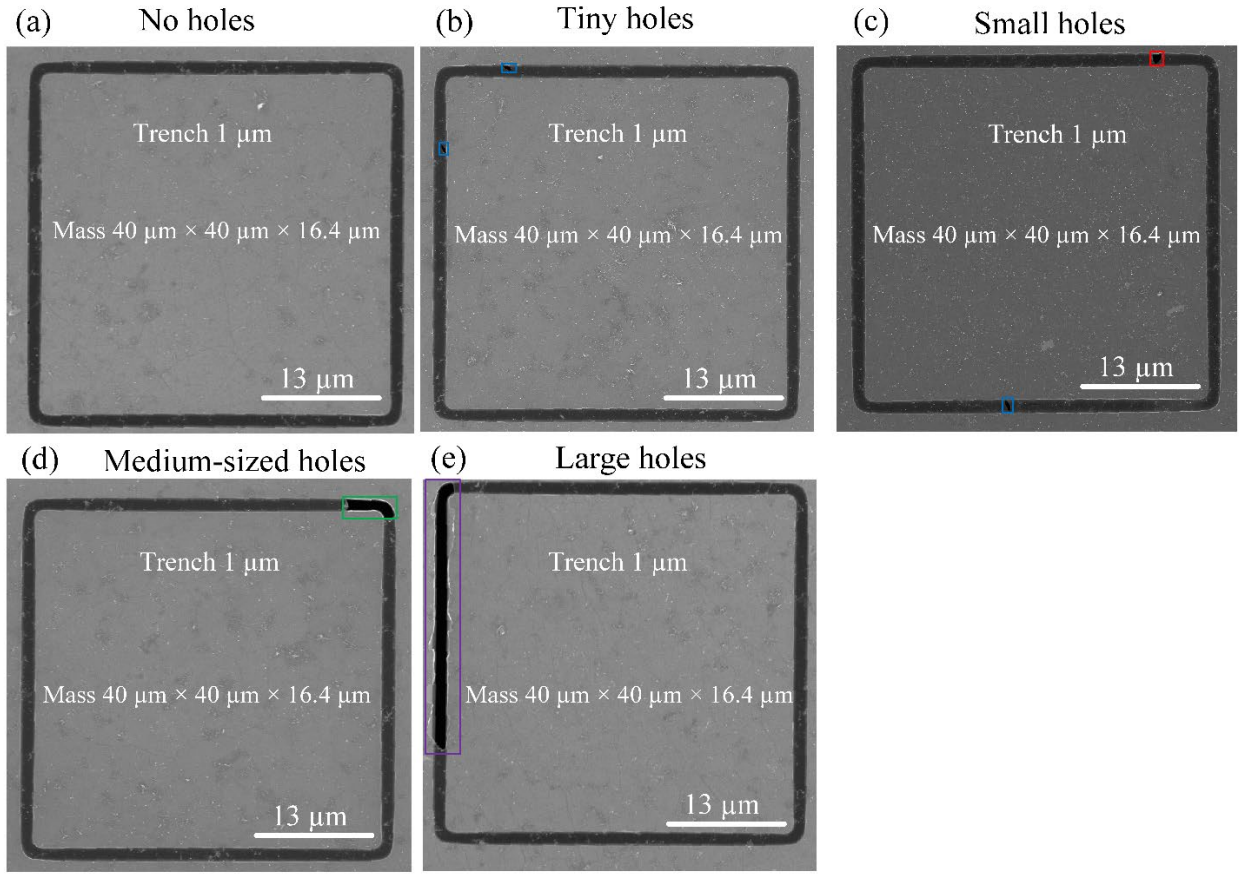

**Fig. 13 SEM images of structures with 1 μm wide trench and 40 μm × 40 μm × 16.4 μm proof masses after annealing at 350 °C for 2 hours. a, b, c, d, e** no holes, tiny holes (blue mark), small holes (red mark), medium-sized holes (green mark) and large holes (purple mark) in suspended graphene membranes, respectively.

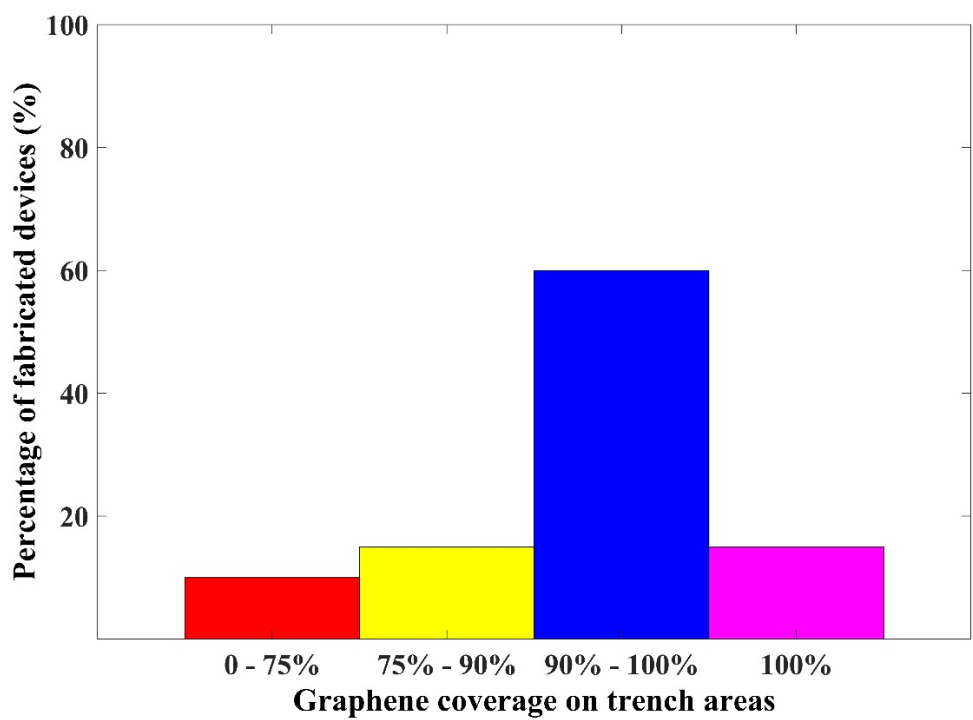

**Fig. 14** Estimated share of fabricated graphene membrane structures that have different percentages of the trench areas covered with double-layer graphene.

## Discussion

For certain applications, it is desirable to obtain graphene membranes with as few polymer residues on the graphene as possible. Such applications include devices for mechanical and electrical characterization of graphene, resonators, high-mobility electronics and gas and biomolecule sensors[60]. During graphene transfer, we used PMMA as a support layer for the graphene, as this allowed easy handling and transfer of the graphene. Even after thorough rinsing with organic solvents such as acetone, PMMA residues (long-chain molecules) remain adhered to the graphene due to the strong dipole interactions between PMMA and chemical groups on graphene[60]. To remove as many of the PMMA residues as possible, we selected some sample chips with released proof masses for annealing at 350 ℃ for 2 hours. During the annealing process, the temperature was first steadily increased from 100 °C to 350 °C in 1 hour and then decreased from 350 °C to 100 °C in 1 hour. The surface of the graphene after annealing was cleaner (Fig. 12) than the surface of graphene without annealing (Fig. 8-12). However, some PMMA residues remained on the surface of the graphene even after annealing (Fig. 3b and c).

To evaluate the graphene quality after annealing, one sample chip was annealed at 350 °C for 1 hour in vacuum and subsequently characterized using Raman spectroscopy (Fig. 15). Fig. 15a shows the Raman spectra of the double-layer graphene at three different positions of the structure. The Raman spectrum shows the typical characteristic peaks of graphene, with the "G peak" occurring at approximately 1600 cm$^{-1}$ (Fig. 15d) and the "2D peak" occurring at approximately 2700 cm$^{-1}$ (Fig. 15e), demonstrating the presence of graphene. The relatively weak "D peak" at positions 1 and 2 occurring at approximately 1359.6 cm$^{-1}$ indicates that the quality of the suspended graphene membranes might decrease to some extent, while a negligible "D peak" at position 3 occurring at approximately 1351 cm$^{-1}$ indicates relatively high quality of graphene (Fig. 15c). The inhomogeneous distribution of stress and doping across the graphene patch might result in correlated variation in the height and position of Raman peaks to some extent. The impact of high-temperature annealing on graphene, such as enhanced hole doping or defects in graphene, has been widely reported on the basis of Raman spectroscopy studies[61–64]. It was also shown that annealing at temperatures below 500°C in vacuum results in a significant decrease in the "D peak" and "2D peak" due to annealing-induced enhanced doping in graphene, and annealing in a vacuum at temperatures of up to 1000°C results in a significant increase in the "2D peak" with a continuous decrease in the "D peak", indicating the partial removal of the defects and restoration of the damaged lattice[63].

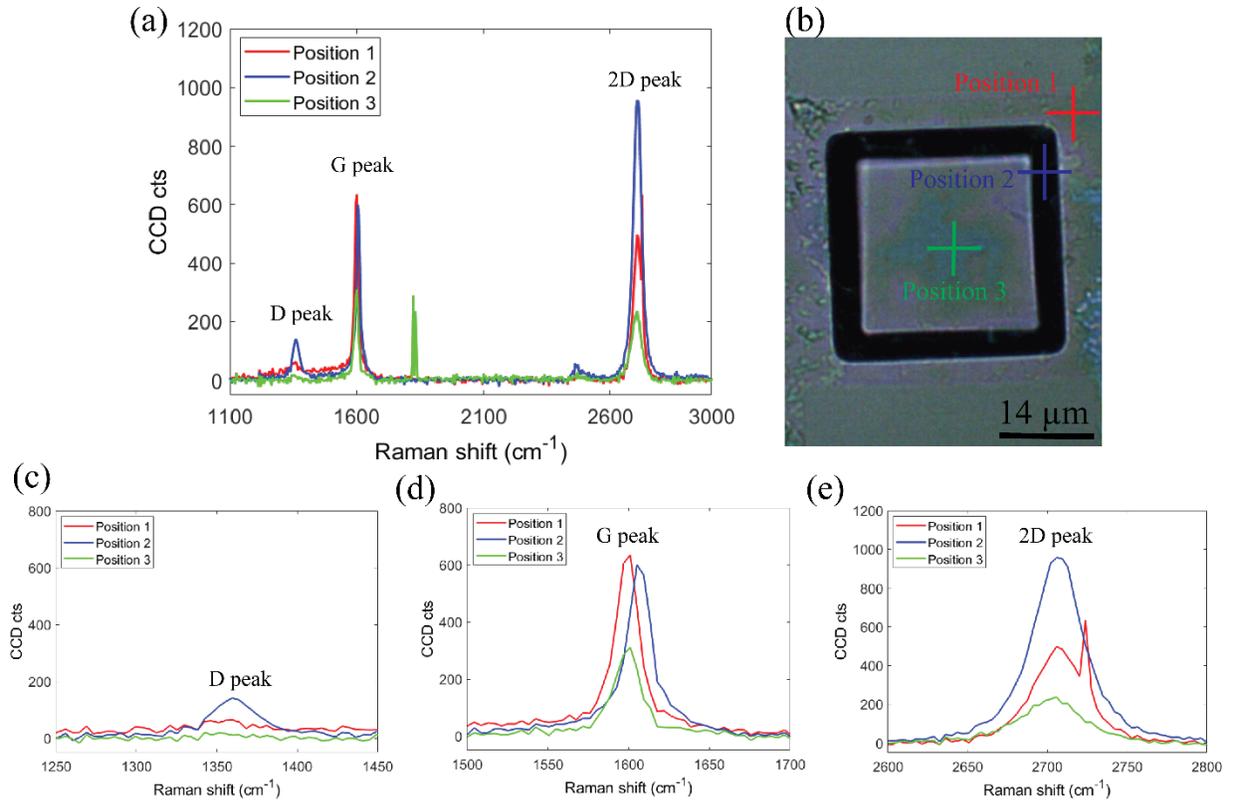

**Fig. 15 Raman spectroscopy of double-layer graphene after annealing. a** Raman spectra of double-layer graphene on three different positions of a structure with 3 μm wide trenches and a 25 μm × 25 μm × 16.4 μm proof mass after annealing at 350 ℃ for 1 hour in vacuum, with "D peaks" occurring at approximately 1359.6 cm$^{-1}$ (position 1), 1359.6 cm$^{-1}$ (position 2) and 1351 cm$^{-1}$ (position 3); "G peaks" occurring at approximately 1601 cm$^{-1}$ (position 1), 1605 cm$^{-1}$ (position 2) and 1601 cm$^{-1}$ (position 3); and "2D peaks" occurring at approximately 2705.6 cm$^{-1}$ (position 1), 2705.6 cm$^{-1}$ (position 2) and 2705.6 cm$^{-1}$ (position 3). **b** Optical microscopy image of the structure characterized in (**a**) with the three different measurement positions. Position 1 (red cross) is on the non-suspended area of double-layer graphene on the substrate; position 2 (blue cross) is on the suspended double-layer graphene membrane; position 3 (green cross) is on the double-layer

graphene on the suspended mass. **c** Magnification of the "D peaks" in (**a**). **d** Magnification of the "G peaks" in (**a**). **e** Magnification of the "2D peaks" in (**a**).

We experimented with several different process flows to fabricate graphene membranes with suspended proof masses and found that wet HF etching was hard to control and often caused graphene displacement, wrinkles or collapse of the graphene due to etching of the $SiO_2$ layer underneath the graphene[65]. In addition, the release process of the proof masses occurring in the liquid environment (liquid HF, etc.) increased the probability of the masses being detached from the suspended graphene membranes due to capillary forces. Only employing HF vapour to etch the BOX layer required a very long time and increased the risks associated with over-etching of the BOX layer. To increase the yields as well as the quality and efficiency of the fabrication process, dry etching followed by vapour HF etching was the preferred approach here.

We also evaluated the possibility of transferring monolayer graphene over the trenches using our baseline process (Fig. 1). However, in this way, it was extremely difficult to obtain suspended monolayer graphene membranes with attached proof masses at high quality and high yield. When using monolayer graphene, our fabrication yield was on the order of 1%, the resulting structures were extremely sensitive, and manual handing was difficult without destroying the structures. We found that the number of holes in a suspended monolayer graphene was extremely high, the size of the holes also increased substantially, and the total coverage area of the suspended monolayer graphene decreased substantially compared to the situation with double-layer graphene membranes. Furthermore, the suspended monolayer graphene completely disappeared over the trenches in many of the fabricated structures. In addition, 100% coverage of monolayer graphene over trenches (meaning that there were no holes in the monolayer graphene) was not achieved in our experiments. We transferred monolayer graphene over trenches on more than 10 chips, and identical results (super low yields) were obtained. Thus, we conclude that double-layer graphene can substantially improve the manufacturing yield of membranes with suspended proof masses compared to monolayer graphene. Double-layer graphene membranes are much stronger than monolayer graphene membranes, which substantially enhances the survival rate of suspended graphene membranes in the entire fabrication process. However, if monolayer graphene membranes with suspended proof masses can be successfully manufactured, for example, by using high-quality CVD monolayer graphene with larger grains that are on the order of hundreds of micrometres in diameter, such membranes would be less stiff and of potential interest for future graphene-based NEMS devices. We also hypothesize that tri-layer graphene or multi-layer graphene would further improve the fabrication yield compared to double-layer graphene, but at the same time, it would most likely increase the membrane stiffness. Increased manufacturing yield and device robustness could be potentially beneficial for large-scale manufacturing of graphene NEMS devices targeted at industrial applications such as accelerometers, gyroscopes and resonators.

In summary, in this paper, we have reported a robust route to transfer and integrate double-layer graphene membranes onto a silicon substrate. The proposed manufacturing process is based on SOI wafer technology and allows the suspension of large silicon proof masses on graphene membranes. Our approach is scalable and highly compatible with silicon NEMS technology and complementary metal oxide semiconductor (CMOS) wafers for the integration of NEMS devices with electronic circuits. The ability of the graphene membranes to withstand AFM indentation forces of up to ~7000 nN without failure indicates that the structures are very robust. Thus, the ability to realize graphene membranes with suspended large proof masses offers interesting opportunities for ultra-miniaturized graphene NEMS devices such as accelerometers, gyroscopes and resonators, with exciting applications in nanoscale robotics, autonomous vehicles, wearable as well as consumer electronics and the internet of things (IoT).

## Materials and methods

The SOI wafer was thermally oxidized to grow a 1.4 μm thick $SiO_2$ layer on both sides of the wafer (Fig. 1f2). A photoresist (PR) layer was spin-coated on the $SiO_2$ surface and patterned to define the trench areas for subsequent etching of the $SiO_2$ and the silicon device layers. Reactive ion etching (RIE) was used to etch the $SiO_2$ layer (Fig. 1f3). The 15 μm thick silicon device layer was etched with deep reactive ion etching (DRIE) to form the trenches and define the proof masses. After silicon trench etching, the remaining PR was removed using oxygen plasma etching (Fig. 1f5 and Fig. 2a). After trench etching, the backside of the SOI wafer was patterned using a PR layer that was spin-coated on the surface of the $SiO_2$ layer on the SOI substrate using lithography with backside alignment (Fig. 1g1 and g2). Then, the $SiO_2$ layer was selectively etched by an RIE etching process (Fig. 1g3). Both the patterned PR and $SiO_2$ layers were used as protection to pattern the silicon handle substrate of the SOI wafer using a DRIE process (Fig. 1g4). The PR residues were then removed by an oxygen plasma etch (Fig. 1g5 and Fig. 2b).

Commercially available CVD monolayer graphene films on copper (Graphenea, Spain) were used. A standard wet transfer approach was employed[66,67], and double-layer graphene was obtained by transferring two graphene monolayers on top of each other (Fig. 1h1-h8). The resulting double-layer graphene was then transferred from the copper substrate to the prefabricated SOI substrate (Fig. 1h9-h11). Then, a poly(methyl methacrylate) (PMMA) solution (AR-P 649.04, ALLRESIST, Germany) was spin-coated on the front side of the first graphene/copper foils at 500 rpm for 5 s followed by 1800 rpm for 30 s and then baked for 5 minutes at 85 °C on a hot plate to evaporate the solvents and cure the PMMA (approximately 200 nm thick) (Fig. 1h2). Then, carbon residues on the backside of the copper foil were removed using $O_2$ plasma etching at low power (80 W) (Fig. 1h3). For wet etching of the copper, the copper foil was placed in a solution of iron(III) chloride hexahydrate ($FeCl_3$), where the copper foil floated on the $FeCl_3$ solution with the graphene side facing away from the liquid. Then, with the help of a silicon carrier wafer, the PMMA/graphene stack without copper (Fig. 1h4) was first transferred onto the surface of deionized (DI) water, then onto a diluted HCl solution and, finally, back to DI water for cleaning, removing the $FeCl_3$ residues and removing chloride residues, respectively. During these transfer processes, it is necessary to keep the PMMA/graphene stack floating on the surface of the liquids and to keep the graphene side on top to ensure that the PMMA covering the graphene is not wetted by the etching solutions. A second graphene on copper foil was used for a second graphene layer transfer (Fig. 1h5). The PMMA/graphene stack floating on the DI water was transferred on the top side of the second graphene/copper foil (Fig. 1h6) and subsequently placed on a hotplate at 45 °C to increase the adhesion between the two graphene layers. Carbon residues on the backside of the copper were removed using $O_2$ plasma etching (Fig. 1h7). A layer of PMMA was spin-coated on the surface of the PMMA/double-layer graphene/copper stack (Fig. 1h8) using process parameters identical to those used for the transfer of the first graphene layer. Then, the same processes were performed to remove the copper substrate (Fig. 1h9) and to transfer the final PMMA/double-layer graphene stack to the pre-patterned SOI substrate (Fig. 1h10). The SOI substrate was then baked at 45 °C on a hotplate for 10 minutes to dry it and to improve the adhesion between the double-layer graphene and the $SiO_2$ surface. Next, the SOI substrate was placed in acetone for 24 hours to remove the PMMA and subsequently placed in isopropanol for 5 minutes to remove the acetone residues. A nitrogen gun was used to gently dry the chip, followed by baking at 45 °C for 10 minutes on a hot plate, which concluded the preparation of the resulting substrates with graphene membranes suspended over trenches (Fig. 1h11 and Fig. 2c1-c5).

To freely suspend the silicon proof masses on the double-layer graphene membranes, RIE dry etching followed by vapour HF etching was used to effectively remove the BOX layer (2 μm thick $SiO_2$) (Fig. 1c), while minimizing the risk of damaging the graphene membranes on the top side of the substrate. Therefore, the chips were attached to the surface of a clean 100 mm diameter silicon carrier wafer by using Kapton tape (Fig. 1i1). To prevent the plasma and the etching gases (such as $CHF_3$, $CF_4$, Ar, $O_2$, $N_2$) from exposing and destroying the graphene, all four sides of the chip were sealed by Kapton tape. Then, an RIE etching process was employed to etch the main part of the BOX layer (Fig. 1i2). To avoid complete removal of the $SiO_2$ layer and subsequent etching and destroying the suspended graphene membranes, only part of the $SiO_2$ layer was etched by carefully tuning the etching time for the $SiO_2$ layer to reach a thickness of the remaining $SiO_2$ layer of approximately 100 nm. HF vapour was then used to continue etching the 100 nm thick $SiO_2$. The vapour HF etching setup (Fig. 2d1) was temperature controlled, and HF vapour was prevented from reaching the front side of the substrate while the $SiO_2$ layer still retained its integrity. A 25% HF solution was placed in the vapour HF chamber, and the temperature was adjusted to 40 ℃. The vapour HF etch rate was calibrated, and the 100 nm thick $SiO_2$ layer was removed in less than 10 minutes, thereby releasing the silicon proof masses and suspending them from the graphene membranes (Fig. 2d). Despite a slight over-etching at the time the $SiO_2$ was removed, the suspended graphene membranes were not destroyed by the short exposure to HF vapour.

Optical microscopy and SEM imaging were used to observe and characterize the morphology of the devices during and after device fabrication. Raman spectroscopy (alpha300 R, WITec) was used to verify the presence and quality of the double-layer graphene of the manufactured devices. For static mechanical characterization, an AFM (Dimension Icon, Bruker) with a cantilever (Olympus AC240TM) and an AFM tip (tip radius = 15 nm) was used to load defined forces at the centre of a proof mass on a graphene membrane to measure the force versus proof mass displacement and the maximum force that the suspended graphene membrane can withstand without rupture. The spring constant of the AFM cantilever was calibrated to be 5.303 N/m. To measure the resonance frequency and quality factor of the structures, we used an LDV (Polytec UHF-120) with a laser spot size on the order of 2.5 μm to detect the amplitude of the thermomechanical noise of the structures in vacuum (~$10^{-5}$ mbar actively pumped vacuum) and the amplitude of structures that were driven by a piezoshaker in air (atmosphere pressure).

## Acknowledgements

The authors would like to thank Cecilia Aronsson, Simon Bleiker, Mikael Bergqvist, Valentin Dubois, Syed Umer Abbas Shah, and Carlos Errando Herranz for fruitful discussions. We acknowledge support through a scholarship from China Scholarship Council, the Starting Grants M&M's (277879) and InteGraDe (307311) as well as Graphene Flagship (785219) from the European Research Council, the Swedish Research Council (GEMS, 2015-05112), the German Federal Ministry for Education and Research (NanoGraM, BMBF, 03XP0006C), and the German Research Foundation (DFG, LE 2440/1-2).

## Conflict of interest

The authors declare that they have no conflict of interest.

## Contributions

X.F., F.N., F.F., A.C.F., A.D.S., and M.C.L. conceived and designed the experiments. A.D.S., S.W., M.Ö. and M.C.L. developed the graphene transfer method. F.F., A.D.S. and S.S contributed to device fabrication. S.W. performed Raman characterization. X.F and S.S.A.A. performed LDV

# References

1. Lee, C., Wei, X., Kysar, J. W. & Hone, J. Measurement of the Elastic Properties and Intrinsic Strength of Monolayer Graphene. *Science* **321**, 385–388 (2008).

2. Bolotin, K. I. *et al.* Ultrahigh electron mobility in suspended graphene. *Solid State Commun.* **146**, 351–355 (2008).

3. Lau, C. N., Bao, W. & Velasco, J. Properties of suspended graphene membranes. *Mater. Today* **15**, 238–245 (2012).

4. Zang, X., Zhou, Q., Chang, J., Liu, Y. & Lin, L. Graphene and carbon nanotube (CNT) in MEMS/NEMS applications. *Microelectron. Eng.* **132**, 192–206 (2015).

5. Qian, Z., Liu, F., Hui, Y., Kar, S. & Rinaldi, M. Graphene as a Massless Electrode for Ultrahigh-Frequency Piezoelectric Nanoelectromechanical Systems. *Nano Lett.* **15**, 4599–4604 (2015).

6. Miao, T., Yeom, S., Wang, P., Standley, B. & Bockrath, M. Graphene Nanoelectromechanical Systems as Stochastic-Frequency Oscillators. *Nano Lett.* **14**, 2982–2987 (2014).

7. Li, P., You, Z. & Cui, T. Graphene cantilever beams for nano switches. *Appl. Phys. Lett.* **101**, 093111 (2012).

8. Castellanos-Gomez, A., Singh, V., van der Zant, H. S. J. & Steele, G. A. Mechanics of freely-suspended ultrathin layered materials. *Ann. Phys.* **527**, 27–44 (2015).

9. Chen, C. & Hone, J. Graphene nanoelectromechanical systems. *Proc. IEEE* **101**, 1766–1779 (2013).

10. Garcia-Sanchez, D. *et al.* Imaging Mechanical Vibrations in Suspended Graphene Sheets. *Nano Lett.* **8**, 1399–1403 (2008).

11. Zande, A. M. van der *et al.* Large-Scale Arrays of Single-Layer Graphene Resonators. *Nano Lett.* **10**, 4869–4873 (2010).

12. Barton, R. A. *et al.* High, Size-Dependent Quality Factor in an Array of Graphene Mechanical Resonators. *Nano Lett.* **11**, 1232–1236 (2011).

13. Shivaraman, S. *et al.* Free-Standing Epitaxial Graphene. *Nano Lett.* **9**, 3100–3105 (2009).

14. Yuasa, Y., Yoshinaka, A., Arie, T. & Akita, S. Visualization of Vibrating Cantilevered Multilayer Graphene Mechanical Oscillator. *Appl. Phys. Express* **4**, 5103 (2011).

15. Bunch, J. S. *et al.* Electromechanical Resonators from Graphene Sheets. *Science* **315**, 490–493 (2007).

16. Eichler, A. *et al.* Nonlinear damping in mechanical resonators made from carbon nanotubes and graphene. *Nat. Nanotechnol.* **6**, 339–342 (2011).

17. Singh, V. *et al.* Probing thermal expansion of graphene and modal dispersion at low-temperature using graphene nanoelectromechanical systems resonators. *Nanotechnology* **21**, 165204 (2010).

18. Singh, V. *et al.* Optomechanical coupling between a multilayer graphene mechanical resonator and a superconducting microwave cavity. *Nat. Nanotechnol.* **9**, 820–824 (2014).

19. Song, X. *et al.* Stamp Transferred Suspended Graphene Mechanical Resonators for Radio Frequency Electrical Readout. *Nano Lett.* **12**, 198–202 (2012).

20. Smith, A. D. *et al.* Electromechanical Piezoresistive Sensing in Suspended Graphene Membranes. *Nano Lett.* **13**, 3237–3242 (2013).

21. Zhu, S.-E., Krishna Ghatkesar, M., Zhang, C. & Janssen, G. C. a. M. Graphene based piezoresistive pressure sensor. *Appl. Phys. Lett.* **102**, 161904 (2013).

22. Dolleman, R. J., Davidovikj, D., Cartamil-Bueno, S. J., van der Zant, H. S. J. & Steeneken, P. G. Graphene Squeeze-Film Pressure Sensors. *Nano Lett.* **16**, 568–571 (2016).

23. Patel, R. N., Mathew, J. P., Borah, A. & Deshmukh, M. M. Low tension graphene drums for electromechanical pressure sensing. *2D Mater.* **3**, 011003 (2016).

24. Davidovikj, D., Scheepers, P. H., van der Zant, H. S. J. & Steeneken, P. G. Static Capacitive Pressure Sensing Using a Single Graphene Drum. *ACS Appl. Mater. Interfaces* **9**, 43205–43210 (2017).

25. Chen, Y.-M. *et al.* Ultra-large suspended graphene as a highly elastic membrane for capacitive pressure sensors. *Nanoscale* **8**, 3555–3564 (2016).

26. Milaninia, K., A. Baldo, M., Reina, A. & Kong, J. All graphene electromechanical switch fabricated by chemical vapor deposition. *Appl. Phys. Lett.* **95**, 183105–183105 (2009).

27. Nagase, M., Hibino, H., Kageshima, H. & Yamaguchi, H. Graphene-Based Nano-Electro-Mechanical Switch with High On/Off Ratio. *Appl. Phys. Express* **6**, 055101 (2013).

28. Sun, J., Muruganathan, M., Kanetake, N. & Mizuta, H. Locally-Actuated Graphene-Based Nano-Electro-Mechanical Switch. *Micromachines* **7**, (2016).

29. Zhou, Q. & Zettl, A. Electrostatic graphene loudspeaker. *Appl. Phys. Lett.* **102**, 223109 (2013).

30. Zhou, Q., Zheng, J., Onishi, S., Crommie, M. F. & Zettl, A. K. Graphene electrostatic microphone and ultrasonic radio. *Proc. Natl. Acad. Sci.* **112**, 8942–8946 (2015).

31. Tian, H. *et al.* Graphene-on-Paper Sound Source Devices. *ACS Nano* **5**, 4878–4885 (2011).

32. Singh, V. & M. Deshmukh, M. Nanoelectromechanics using graphene. *Curr. Sci.* **107**, 437–446 (2014).

33. Conley, H., Lavrik, N. V., Prasai, D. & Bolotin, K. I. Graphene Bimetallic-like Cantilevers: Probing Graphene/Substrate Interactions. *Nano Lett.* **11**, 4748–4752 (2011).

34. Kim, H. J., Choi, J., Nam, S. & King, W. P. Batch Fabrication of Transfer-Free Graphene-Coated Microcantilevers. *IEEE Sens. J.* **15**, 2717–2718 (2015).

35. Shim, W. *et al.* Multifunctional cantilever-free scanning probe arrays coated with multilayer graphene. *Proc. Natl. Acad. Sci. U. S. A.* **109**, 18312–18317 (2012).

36. Rasuli, R., zad, A. I. & Ahadian, M. M. Mechanical properties of graphene cantilever from atomic force microscopy and density functional theory. *Nanotechnology* **21**, 185503 (2010).

37. Reserbat-Plantey, A., Marty, L., Arcizet, O., Bendiab, N. & Bouchiat, V. A local optical probe for measuring motion and stress in a nanoelectromechanical system. *Nat. Nanotechnol.* **7**, 151–155 (2012).

38. Schwarz, C. *et al.* Deviation from the Normal Mode Expansion in a Coupled Graphene-Nanomechanical System. *Phys. Rev. Appl.* **6**, (2016).

39. Wang, Q., Hong, W. & Dong, L. Graphene "microdrums" on a freestanding perforated thin membrane for high sensitivity MEMS pressure sensors. *Nanoscale* **8**, 7663–7671 (2016).

40. Chen, C. *et al.* Performance of monolayer graphene nanomechanical resonators with electrical readout. *Nat. Nanotechnol.* **4**, 861–867 (2009).

41. Stolyarova, E. *et al.* Observation of Graphene Bubbles and Effective Mass Transport under Graphene Films. *Nano Lett.* **9**, 332–337 (2009).

42. Lee, S. *et al.* Electrically integrated SU-8 clamped graphene drum resonators for strain engineering. *Appl. Phys. Lett.* **102**, 153101 (2013).

43. Maurand, R. *et al.* Fabrication of ballistic suspended graphene with local-gating. *Carbon* **79**, 486–492 (2014).

44. Tombros, N. *et al.* Large yield production of high mobility freely suspended graphene electronic devices on a polydimethylglutarimide based organic polymer. *J. Appl. Phys.* **109**, 093702 (2011).

45. Matsui, K. *et al.* Mechanical properties of few layer graphene cantilever. in *2014 IEEE 27th International Conference on Micro Electro Mechanical Systems (MEMS)* 1087–1090 (IEEE, 2014). doi:10.1109/MEMSYS.2014.6765834

46. Blees, M. K. *et al.* Graphene kirigami. *Nature* **524**, 204–207 (2015).

47. Hurst, A. M., Lee, S., Cha, W. & Hone, J. A graphene accelerometer. in *2015 28th IEEE International Conference on Micro Electro Mechanical Systems (MEMS)* 865–868 (2015). doi:10.1109/MEMSYS.2015.7051096

48. Bunch, J. S. & Dunn, M. L. Adhesion mechanics of graphene membranes. *Solid State Commun.* **152**, 1359–1364 (2012).

49. Koenig, S. P., Boddeti, N. G., Dunn, M. L. & Bunch, J. S. Ultrastrong adhesion of graphene membranes. *Nat. Nanotechnol.* **6**, 543–546 (2011).

50. Lu, Z. & Dunn, M. L. van der Waals adhesion of graphene membranes. *J. Appl. Phys.* **107**, 044301 (2010).

51. Ferrari, A. C. *et al.* Raman spectrum of graphene and graphene layers. *Phys. Rev. Lett.* **97**, 187401 (2006).

52. Malard, L. M., Pimenta, M. A., Dresselhaus, G. & Dresselhaus, M. S. Raman spectroscopy in graphene. *Phys. Rep.* **473**, 51–87 (2009).

53. Ni, Z. H. *et al.* Graphene Thickness Determination Using Reflection and Contrast Spectroscopy. *Nano Lett.* **7**, 2758–2763 (2007).

54. Bunch, J. S. *et al.* Impermeable Atomic Membranes from Graphene Sheets. *Nano Lett.* **8**, 2458–2462 (2008).

55. Barton, R. A., Parpia, J. & Craighead, H. G. Fabrication and performance of graphene nanoelectromechanical systems. *J. Vac. Sci. Technol. B* **29**, 050801 (2011).

56. Bhiladvala, R. B. & Wang, Z. J. Effect of fluids on the Q factor and resonance frequency of oscillating micrometer and nanometer scale beams. *Phys. Rev. E Stat. Nonlin. Soft Matter Phys.* **69**, 036307 (2004).

57. Vishwakarma, S. D. *et al.* Size modulated transition in the fluid–structure interaction losses in nano mechanical beam resonators. *J. Appl. Phys.* **119**, 194303 (2016).

58. López-Polín, G. *et al.* Increasing the elastic modulus of graphene by controlled defect creation. *Nat. Phys.* **11**, 26–31 (2015).

59. Lee, G.-H. *et al.* High-Strength Chemical-Vapor–Deposited Graphene and Grain Boundaries. *Science* **340**, 1073–1076 (2013).

60. Lin, Y.-C. *et al.* Clean Transfer of Graphene for Isolation and Suspension. *ACS Nano* **5**, 2362–2368 (2011).

61. Kaplas, T. *et al.* Effect of High-Temperature Annealing on Graphene with Nickel Contacts. *Condens. Matter* **4**, 21 (2019).

62. Song, J., Ko, T. Y. & Ryu, S. Raman Spectroscopy Study of Annealing-Induced Effects on Graphene Prepared by Micromechanical Exfoliation. *Bull. Korean Chem. Soc.* **31**, 2679–2682 (2010).

63. Zion, E. *et al.* Effect of annealing on Raman spectra of monolayer graphene samples gradually disordered by ion irradiation. *J. Appl. Phys.* **121**, 114301 (2017).

64. Son, J., Choi, M., Hong, J. & Yang, I.-S. Raman study on the effects of annealing atmosphere of patterned graphene: Effects of different annealing atmosphere. *J. Raman Spectrosc.* **49**, 183–188 (2018).

65. Aydin, O. I., Hallam, T., Thomassin, J. L., Mouis, M. & Duesberg, G. S. Interface and strain effects on the fabrication of suspended CVD graphene devices. *Solid-State Electron.* **108**, 75–83 (2015).

66. Suk, J. W. *et al.* Transfer of CVD-Grown Monolayer Graphene onto Arbitrary Substrates. *ACS Nano* **5**, 6916–6924 (2011).

67. Li, X. *et al.* Transfer of Large-Area Graphene Films for High-Performance Transparent Conductive Electrodes. *Nano Lett.* **9**, 4359–4363 (2009).